\documentclass[12pt,eqsecnum]{article}
\usepackage[dvips]{graphicx}
\usepackage{amssymb}
\usepackage{amsmath}
\usepackage{ulem}
\usepackage[dvipsnames]{xcolor}
\usepackage{soul}
\usepackage{ascmac}
\usepackage{cite}
\usepackage{url}
\makeatletter

\@addtoreset{equation}{section}
\makeatletter

\newcommand{\D}{{\rm d}}

\newtheorem{The}{Theorem}

\newcommand{\dalm}{\kern1pt\vbox{\hrule height 0.9pt\hbox{\vrule width
0.9pt\hskip 2.5pt\vbox{\vskip 5.5pt}\hskip 3pt\vrule width 0.3pt}\hrule height
0.3pt}\kern1pt}

\begin{document}

\begin{titlepage}
\vfill
\begin{flushright}
\today
\end{flushright}

\vfill
%\vskip 1.0cm
\begin{center}
\baselineskip=16pt
{\Large\bf 
Simple traversable wormholes violating energy conditions only near the Planck scale}
\vskip 0.5cm
{\large {\sl }}
\vskip 10.mm
{\bf Hideki Maeda} \\

\vskip 1cm
{
Department of Electronics and Information Engineering, Hokkai-Gakuen University, Sapporo 062-8605, Japan.\\
\texttt{h-maeda@hgu.jp}

}
\vspace{6pt}
%\today
\end{center}
\vskip 0.2in
\par
\begin{center}
{\bf Abstract}
\end{center}
\begin{quote}
We present a static and axisymmetric traversable wormhole spacetime with vanishing Arnowitt-Deser-Misner (ADM) mass which is characterized by a length parameter $l$ and a deformation parameter $a$ and reduces to the massless Kerr vacuum wormhole as $l\to 0$.
The spacetime is analytic everywhere and regularizes a ring-like conical singularity of the massless Kerr wormhole by virtue of a localized exotic matter which violates the standard energy conditions only near the wormhole throat.
In the spherically symmetric case ($a=0$), the areal radius of the wormhole throat is exactly $l$ and all the standard energy conditions are respected outside the proper radial distance approximately $1.60l$ from the throat.
While the curvature at the throat is beyond the Planck scale if $l$ is identical to the Planck length $l_{\rm p}$, our wormhole may be a semi-classical model for $l\simeq 10l_{\rm p}$.
With $l=10l_{\rm p}$, the total amount of the negative energy supporting this wormhole is only $E\simeq -26.5m_{\rm p}c^2$, which is the rest mass energy of about $-5.77\times 10^{-4}{\rm g}$.
It is shown that the geodesic behavior on the equatorial plane does not qualitatively change by the localization of an exotic matter.
\vfill
% \hrule width 5.cm
\vskip 2.mm
\end{quote}
\end{titlepage}

%<<<<<<<<<<<<< PACS NUMBER >>>>>>>>>>>>>>>%
%\pacs{
%04.20.]q Classical general relativity
%04.20.Gz Spacetime topology, causal structure, spinor structure
%} 

% CECS-PHY-13/09

%\maketitle
%\section{}
%\subsection{}

\tableofcontents

\newpage

%======================================%
%<<<<<<<<<<<< SECTION 1 >>>>>>>>>>>>>>%
%======================================%
\section{Introduction}
\begin{figure}[htbp]
\begin{center}
%\rotatebox{-90}{
\includegraphics[width=0.4\linewidth]{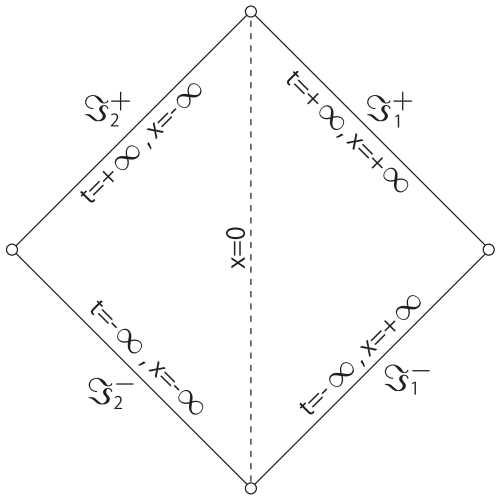}
%\subfigure[]{\includegraphics[width=0.7\linewidth]{Roberts-lambda1.eps}}
%\subfigure[]{\includegraphics[width=0.7\linewidth]{Roberts-lambda2.eps}}
%}
\caption{\label{wormhole} A Penrose diagram of the static wormhole spacetime (\ref{semi-vacuum}).
The future (past) null infinities $\Im^{+(-)}_1$ and $\Im^{+(-)}_2$ are distinct.
}
\end{center}
\end{figure}

Wormhole is a configuration of spacetimes connecting distinct non-timelike infinities.
In particular, a wormhole that admits a causal curve connecting such infinities are referred to as a {\it traversable wormhole}.
According to this definition, the maximally extended Schwarzschild spacetime is a dynamical wormhole but it is not traversable~\cite{Einstein:1935tc,Fuller:1962zza}.
A typical Penrose diagram of a traversable wormhole is shown in Fig.~\ref{wormhole}.
Traversable wormhole is one of the most interesting research objects in gravitation physics because it has potential applications to realize an apparently superluminal spacetime shortcut~\cite{Morris:1988cz,Lobo:2017oab} or to create a time machine~\cite{Morris:1988tu,Kim:1991mc,Visser:1992tx}.
(See~\cite{Visser:1995cc} for a textbook and~\cite{Lobo:2007zb} for a review.)
%------------<fig>---------------------------
%--------------<fig>-----------------------

However, in order to realize a traversable wormhole, we need an exotic matter, of which energy-momentum tensor $T_{\mu\nu}$ in the Einstein equations $G_{\mu\nu}=8\pi GT_{\mu\nu}$ (with units such that $c=1$) violates all the standard energy conditions.
The standard energy conditions for $T_{\mu\nu}$ are as follows~\cite{Hawking:1973uf,Maeda:2018hqu}:
\begin{itemize}
\item {Null} energy condition (NEC): $T_{\mu\nu} k^\mu k^\nu\ge 0$ for any null vector $k^\mu$.
\item {Weak} energy condition (WEC): $T_{\mu\nu} v^\mu v^\nu\ge 0$ for any timelike vector $v^\mu$.
\item {Dominant} energy condition (DEC): $T_{\mu\nu} v^\mu v^\nu\ge 0$ and $J_\mu J^\mu\le 0$ hold for any timelike vector $v^\mu$, where $J^\mu:=-T^\mu_{\phantom{\mu}\nu}v^\nu$. 
\item {Strong} energy condition (SEC): $\left(T_{\mu\nu}-\frac{1}{2}Tg_{\mu\nu}\right) v^\mu v^\nu\ge 0$ for any timelike vector $v^\mu$.
\end{itemize}
The NEC (WEC) means non-negativity of the energy density measured by any null (causal) observer.
The condition $J_\mu J^\mu\le 0$ in the DEC means that the energy current associated with a causal observer is absent or does not propagate faster than the speed of light.
The SEC is equivalent to the timelike convergence condition (TCC) $R_{\mu\nu}v^\mu v^\nu\ge 0$ in general relativity in the absence of the cosmological constant $\Lambda$, which implies that gravity is essentially attractive.
The TCC includes the null convergence condition (NCC) $R_{\mu\nu}k^\mu k^\nu\ge 0$ as a limiting case.

Since the relations of the energy conditions are DEC$\subset$WEC$\subset$NEC and SEC$\subset$NEC, violation of the NEC means that all the standard energy conditions are violated.
In fact, by the following topological censorship theorem~\cite{Friedman:1993ty}, traversable wormholes cannot be realized in general relativity even under a condition weaker than the NEC.
(See also~\cite{Galloway:1995hidden}.)
%----------------------- lemma ------------------------------%
\begin{The}
\label{The:topological}(Topological censorship theorem)
If an asymptotically flat, globally hyperbolic spacetime satisfies the averaged null convergence condition (ANCC), then every causal curve from a past null infinity $\Im^{-}$ to a future null infinity $\Im^{+}$ can be continuously deformed to a topologically trivial curve from $\Im^{-}$ to $\Im^{+}$ near infinity.
\end{The}
%----------------------- lemma ------------------------------%
For example, a causal curve from $\Im^{-}_1$ to $\Im^{+}_2$ in Fig.~\ref{wormhole} cannot be continuously deformed to a curve from $\Im^{-}_1$ to $\Im^{+}_1$, so that the ANCC is violated in this spacetime.
Here the ANCC is defined by $\int R_{\mu\nu}k^\mu k^\nu\D \lambda\ge 0$ along every inextendible null geodesic parametrized by an affine parameter $\lambda$ and equivalent to the averaged null energy condition (ANEC) $\int T_{\mu\nu}k^\mu k^\nu\D \lambda\ge 0$ in general relativity.
Due to the relation NEC$\subset$ANEC, violation of the NEC is not enough for a traversable wormhole and there must be at least one geodesic along which $\int R_{\mu\nu}k^\mu k^\nu\D \lambda$ is negative.

Nevertheless, Visser et al. presented explicit examples of traversable wormholes in which the magnitude of the integral $\int R_{\mu\nu}k^\mu k^\nu\D \lambda$ can be arbitrarily small~\cite{Visser:2003yf}.
Their wormholes are constructed by gluing two spacetimes at a timelike hypersurface $\Sigma$ with vanishing extrinsic curvature, so that the spacetimes are $C^2$ at $\Sigma$.
It is still not clear if the magnitude of $\int R_{\mu\nu}k^\mu k^\nu\D \lambda$ can be arbitrarily small for a wormhole spacetime which is analytic everywhere.

Needless to say, in addition to the problem of the energy conditions, dynamical stability is the most important and difficult problem to achieve a realistic traversable wormhole.
Unfortunately, in the case where a spacetime is analytic everywhere, few dynamically stable static wormholes are known not only in general relativity but also in a wide class of generalized theories of gravity.
(See~\cite{Bronnikov:2013coa} for one of the few positive results.)
However, it has been reported that a class of scalar-tensor theories called beyond-Horndeski theories~\cite{Zumalacarregui:2013pma,Gleyzes:2014dya} could admit dynamically stable static wormholes~\cite{Franciolini:2018aad,Mironov:2018uou}.
(See review papers~\cite{Kobayashi:2019hrl,Frusciante:2019xia,Kase:2018aps} for cosmologically viable models in beyond-Horndeski theories.)
Since the instability of a wormhole is associated with the violation of the energy conditions, a dynamically stable configuration could be realized if the region of violation is sufficiently small.
In this paper, we present a simple example of such traversable wormhole spacetimes.

The organization of the present paper is as follows.
In Sec.~\ref{sec:model}, we will present our spherically symmetric model of traversable wormholes with a localized exotic matter and then generalize it to be axisymmetric which regularizes the massless Kerr vacuum wormhole with a ring-like conical singularity.
In Sec.~\ref{app:Petrov}, we will study geometric properties of our wormhole spacetimes.
Concluding remarks and future prospects will be given in the final section.
Our conventions for curvature tensors are $[\nabla _\rho ,\nabla_\sigma]V^\mu ={R^\mu }_{\nu\rho\sigma}V^\nu$ and $R_{\mu \nu }={R^\rho }_{\mu \rho \nu }$.
The signature of the Minkowski spacetime is $(-,+,+,+)$, and Greek indices run over all spacetime indices.
We adopt units such that $c=1$ and the gravitational constant $G$ is shown explicitly.

%======================================%
%<<<<<<<<<<<< SECTION 1 >>>>>>>>>>>>>>%
%======================================%
\section{Static wormholes with a localized exotic matter}
\label{sec:model}

In this section, we present our model of static traversable wormholes with an exotic matter localized near the wormhole throat.
The metric is given in the coordinates $\{t,x,\theta,\phi\}$ in the following diagonal form:
\begin{align}
\D s^2=g_{tt}(x,\theta)\D t^2+g_{xx}(x,\theta)\D x^2+g_{\theta\theta}(x,\theta)\D \theta^2+g_{\phi\phi}(x,\theta)\D \phi^2,
\end{align}
where $g_{tt}<0$ and $g_{xx},g_{\theta\theta},g_{\phi\phi}>0$ hold everywhere.
The energy-momentum tensor in this spacetime is given in a diagonal form as $T^\mu_{~\nu}:=(8\pi G)^{-1}G^\mu_{~\nu}={\rm diag}(-\mu,p_1,p_2,p_3)$.
Then, with the orthonormal basis one-forms $\{{E}_\mu^{(a)}\}~(a=0,1,2,3)$ in the local Lorentz frame
\begin{align}
\begin{aligned}
&E_\mu^{(0)}\D x^\mu=\sqrt{-g_{tt}}\D t,\qquad E_\mu^{(1)}\D x^\mu=\sqrt{g_{xx}}\D x,\\
&E_\mu^{(2)}\D x^\mu=\sqrt{g_{\theta\theta}}\D \theta,\qquad E_\mu^{(3)}\D x^\mu=\sqrt{g_{\phi\phi}}\D \phi,
\end{aligned}
\end{align}
which satisfy ${E}^\mu_{(a)}{E}_{(b)\mu}=\mbox{diag}(-1,1,1,1)$, we obtain orthonormal components of ${T}_{\mu\nu}$ as
\begin{equation} 
\label{T-typeI}
{T}^{(a)(b)}:={T}^{\mu\nu} {E}_\mu^{(a)} {E}_\nu^{(b)}=\mbox{diag}(\mu,p_1,p_2,p_3).
\end{equation}
This is the type I energy-momentum tensor in the Hawking-Ellis classification~\cite{Hawking:1973uf,Maeda:2018hqu} and the equivalent expressions to the standard energy conditions for this type of matter are
\begin{align}
\mbox{NEC}:&~~\mu+p_i\ge 0,\label{NEC}\\
\mbox{WEC}:&~~\mu\ge 0\mbox{~in addition to NEC},\label{WEC}\\
\mbox{DEC}:&~~\mu-p_i\ge 0\mbox{~in addition to WEC},\label{DEC}\\
\mbox{SEC}:&~~\mu+\mbox{$\sum_{j=1}^{3}$}p_j\ge 0\mbox{~in addition to NEC}\label{SEC}
\end{align}
for all $i(=1,2,3)$.

\subsection{Spherically symmetric model}
\label{sec:spherical}
\begin{figure}[htbp]
\begin{center}
%\rotatebox{-90}{
\includegraphics[width=0.6\linewidth]{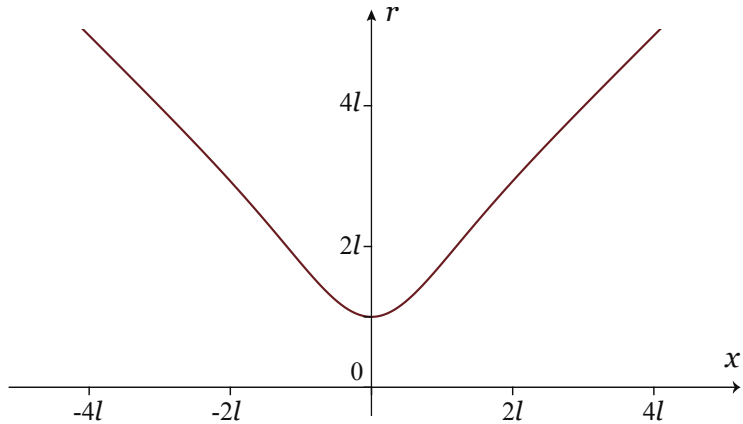}
%\subfigure[]{\includegraphics[width=0.7\linewidth]{Roberts-lambda1.eps}}
%\subfigure[]{\includegraphics[width=0.7\linewidth]{Roberts-lambda2.eps}}
%}
\caption{\label{Simplewormhole} The areal radius $r(x)$ of the spacetime (\ref{semi-vacuum}). $r''>0$ holds in the region $-1.20l\lesssim x\lesssim 1.20l$. 
}
\end{center}
\end{figure}

Our static and spherically symmetric model is given by 
\begin{align}
\label{semi-vacuum}
\begin{aligned}
&\D s^2=-\D t^2+\D x^2+r(x)^2(\D \theta^2+\sin^2\theta\D\phi^2),\\
&r(x)=l+x\tanh(x/l)
\end{aligned}
\end{align} 
defined in the domain $x\in(-\infty,\infty)$, where $l$ is a parameter with the dimension of length.
The shape of the areal radius $r(x)$ is shown in Fig.~\ref{Simplewormhole}.
$r(x)$ is symmetric with respect to $x=0$ and has a single local minimum $r(0)=l$, around which it behaves as
\begin{align}
\lim_{x\to 0}\frac{r(x)}{l}=&1+(x/l)^2+O((x/l)^4).
\end{align} 
By the theorem in~\cite{Lobo:2020ffi}, all the standard energy conditions are violated in a region where $r''>0$ holds, where a prime denotes differentiation with respect to $x$. (See Proposition 4 in~\cite{Maeda:2021ukk} for a generalization.)
In our model (\ref{semi-vacuum}), this sufficient condition $r''>0$ holds in the region $-1.20l\lesssim x\lesssim 1.20l$.
%------------<fig>---------------------------
%--------------<fig>-----------------------

The spacetime (\ref{semi-vacuum}) converges to Minkowski in the limit $l\to 0$ in a non-uniform manner.
The areal radius $r(x)$ is analytic everywhere and obeys
\begin{align}
\lim_{x/l\to \pm\infty}\frac{r(x)}{x}=&\pm 1+\frac{l}{x}\mp 2e^{\mp 2x/l}+O(e^{\mp 4x/l}),\label{expand-r}
\end{align} 
which is not the Taylor series.
%~\footnote{We note that all the standard energy conditions are violated in the asymptotically flat region in the case of $r(x)=\sqrt{l^2+x^2\tanh^2(x/l)}$.}.
Equation~(\ref{expand-r}) shows that the spacetime (\ref{semi-vacuum}) is asymptotically flat as $x\to \pm \infty$ with the vanishing ADM mass~\cite{ChruscielLecture}.
As a result, the spacetime (\ref{semi-vacuum}) represents an asymptotically flat static traversable wormhole with a single wormhole throat at $x=0$, of which Penrose diagram is given by Fig.~\ref{wormhole}.

A fascinating property of the wormhole (\ref{semi-vacuum}) is that the standard energy conditions are violated only around the throat at $x=0$.
This is a sharp difference from the massless Ellis-Bronnikov wormhole~\cite{Ellis1973,Bronnikov1973}, described by the spacetime (\ref{semi-vacuum}) with $r(x)=\sqrt{x^2+l^2}$.
In this case, all the standard energy conditions are violated everywhere because of $r''=l^2/(x^2+l^2)^{3/2}>0$.

To simplify the analysis, we introduce dimensionless coordinates $w:=x/l$ and $T:=t/l$, with which our spacetime (\ref{semi-vacuum}) is written as
\begin{align}
\label{semi-vacuum-w}
\begin{aligned}
&\D s^2=l^2\left\{-\D T^2+\D w^2+\sigma(w)^2(\D \theta^2+\sin^2\theta\D\phi^2)\right\},\\
&\sigma(w):=r(x(w))/l=1+w\tanh w.
\end{aligned} 
\end{align} 
The corresponding energy-momentum tensor is diagonal such that $T^\mu_{~\nu}:=(8\pi G)^{-1}G^\mu_{~\nu}={\rm diag}(-\mu,p_1,p_2,p_2)$, where
\begin{align}
\mu=&-\frac{2\sigma\sigma''+{\sigma'}^2-1}{8\pi Gl^2\sigma^2}=\frac{(5w\tanh w+3)(w\tanh w - 1)-w^2}{8\pi Gl^2\sigma(w)^2\cosh^2 w},\label{rho}\\
p_1=&\frac{{\sigma'}^2-1}{8\pi Gl^2\sigma^2}=\frac{w^2-(w\tanh w-1)^2}{8\pi Gl^2\sigma(w)^2\cosh^2 w},\\
p_2=&\frac{\sigma''}{8\pi Gl^2\sigma}=\frac{2(1-w\tanh w)}{8\pi Gl^2\sigma(w)\cosh^2 w}.\label{p2}
\end{align} 
Hence, the spacetime (\ref{semi-vacuum}), or equivalently the spacetime (\ref{semi-vacuum-w}), is an exact solution in general relativity with an anisotropic fluid, of which energy-momentum tensor is given by 
\begin{align}
&{T}_{\mu\nu}=(\mu+p_2)u_\mu u_\nu+(p_1-p_2)s_\mu s_\nu +p_2g_{\mu\nu},\label{T-fluid}\\
&u^\mu\frac{\partial}{\partial x^\mu}=l^{-1}\frac{\partial}{\partial T},\qquad s^\mu\frac{\partial}{\partial x^\mu}=l^{-1}\frac{\partial}{\partial w},
\end{align} 
where $\mu$, $p_1$, and $p_2$ are the energy density, radial pressure, and tangential pressure of the fluid, respectively.
Here $u^\mu$ and $s^\mu$ are unit timelike and spacelike vectors, respectively, satisfying $u_\mu u^\mu=-1$, $s_\mu s^\mu=1$, and $u_\mu s^\mu=0$.

Equations~(\ref{rho})--(\ref{p2}) give $\mu+p_1+2p_2=0$ and 
\begin{align}
&\mu+p_1=-\frac{\sigma''}{4\pi Gl^2\sigma}=\frac{w\tanh w-1}{2\pi Gl^2\sigma(w)\cosh^2 w},\\
&\mu+p_2=\frac12(\mu-p_1) \nonumber \\
&~~~~~~~~=-\frac{\sigma\sigma''+{\sigma'}^2-1}{8\pi Gl^2\sigma^2}=\frac{(3w\tanh w+1)(w\tanh w - 1)-w^2}{8\pi Gl^2\sigma(w)^2\cosh^2 w},\label{rho+p1}\\
&\mu-p_2=-\frac{3\sigma\sigma''+{\sigma'}^2-1}{8\pi Gl^2\sigma^2}=\frac{(7w\tanh w+5)(w\tanh w - 1)-w^2}{8\pi Gl^2\sigma(w)^2\cosh^2 w},\label{rho+p1+2p2}
\end{align} 
which show
\begin{align}
\mu>0~~&\leftrightarrow~~|w|\gtrsim 1.37,\\
\mu+p_1>0~~&\leftrightarrow~~|w|\gtrsim 1.20,\\
\mu+p_2>0, \mu-p_1>0~~&\leftrightarrow~~|w|\gtrsim 1.60,\\
\mu-p_2>0~~&\leftrightarrow~~|w|\gtrsim 1.31.
\end{align} 
Thus, according to Eqs.~(\ref{NEC})--(\ref{SEC}), all the standard energy conditions are violated (respected) in the region $|x|\lesssim(\gtrsim) 1.60l$ in the spacetime (\ref{semi-vacuum}).

%\begin{align}
%&-\mu|_{x=0}=-\frac{3}{8\pi Gl^2}\simeq -\frac{0.119}{Gl^2}
%\end{align}

\begin{figure}[htbp]
\begin{center}
%\rotatebox{-90}{
\includegraphics[width=0.5\linewidth]{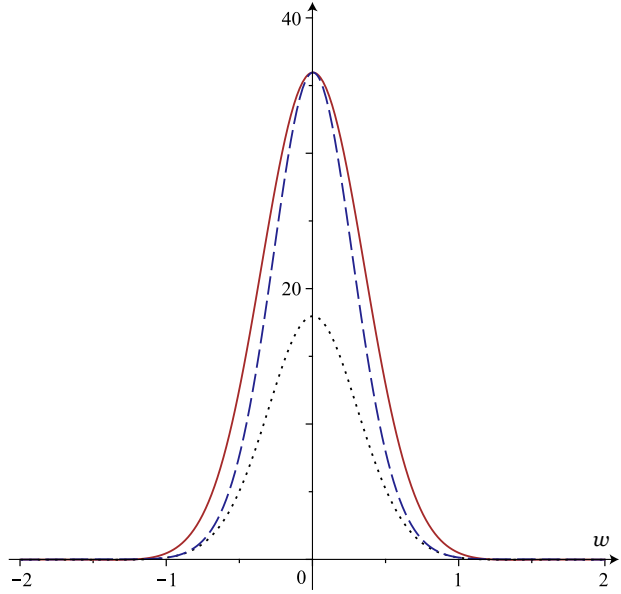}
%\subfigure[]{\includegraphics[width=0.7\linewidth]{Roberts-lambda1.eps}}
%\subfigure[]{\includegraphics[width=0.7\linewidth]{Roberts-lambda2.eps}}
%}
\caption{\label{Curvature-Wormhole} Curvature invariants $l^4R^2$ (solid), $l^4R_{\mu\nu}R^{\mu\nu}$ (dotted), and $l^4R_{\mu\nu\rho\sigma}R^{\mu\nu\rho\sigma}$ (dashed) of the wormhole spacetime (\ref{semi-vacuum}) as functions of $w(=x/l)$.
}
\end{center}
\end{figure}

As shown in Fig.~\ref{Curvature-Wormhole}, curvature invariants of the spacetime (\ref{semi-vacuum}) take the maximum values at the throat ($x=0$) as
\begin{align}
&R|_{x=0}=-\frac{6}{l^2}=-\frac{6l_{\rm p}^2}{l^2}{\cal R}_p,\\
&R_{\mu\nu}R^{\mu\nu}|_{x=0}=\frac{18}{l^4}=\frac{18l_{\rm p}^4}{l^4}{\cal R}_p^2,\\
&R_{\mu\nu\rho\sigma}R^{\mu\nu\rho\sigma}|_{x=0}=\frac{36}{l^4}=\frac{36l_{\rm p}^4}{l^4}{\cal R}_p^2, \label{C-invariant}
\end{align}
where $l_{\rm p}:=\sqrt{\hbar G/c^3}(\simeq 1.616\times 10^{-35}{\rm m})$ is the Planck length and ${\cal R}_{\rm p}:=c^3/(\hbar G)(=l_{\rm p}^{-2})$ is the Planck curvature.
If the parameter $l$ is identical to the Planck length $l=l_{\rm p}$, the curvature at the wormhole throat is larger than the Planck curvature and therefore full quantum gravity should be necessary to describe there.
With $l\simeq 10l_{\rm p}$, the curvature is less than the Planck scale and then the wormhole (\ref{semi-vacuum}) may be justified as a semi-classical model.
%------------<fig>---------------------------
%--------------<fig>-----------------------

The total amount of the negative energy on a spacelike hypersurface with constant $t$ is computed to give
\begin{align}
E(1.37l)\simeq -2.65lG^{-1}=:E_{-}(l), \label{negative-E}
%=-2.65lG^{-1}c^4
\end{align} 
where 
\begin{align}
E(x):=4\pi\int_{-x}^{x}\mu({\bar x}) r({\bar x})^2\D {\bar x}. \label{def-E}
\end{align} 
If $l=10l_{\rm p}$, this amount is $E_{-}(10l_{\rm p})=-26.5m_{\rm p}c^2$, where $m_{\rm p}:=\sqrt{c\hbar/G}(\simeq 2.176\times 10^{-5}{\rm g})$ is the Planck mass and we have written the speed of light $c$ explicitly.
This is the rest mass energy of about $-5.77\times 10^{-4}{\rm g}$.
For a ``humanly traversable'' wormhole with $l=1{\rm m}$, we obtain $E_{-}(1{\rm m})/c^2\simeq -3.57\times 10^{27}{\rm kg}$, which is about $-1.88$ times as large as Jupiter's mass.
Although the region with negative energy density is finite, the total amount of the energy on a spacelike hypersurface with constant $t$ in the spacetime (\ref{semi-vacuum}) is still negative as $\lim_{x\to\infty}E(x)\simeq -1.79lG^{-1}$.
%\begin{align}
%E\simeq &4\pi\int_{-1.37l}^{1.37l}\mu r(x)^2\D x \nonumber \\
%=&8\pi l^3\int_{0}^{1.37}\biggl(-\frac{1}{8\pi G}G^T_{~T}\biggl) \sigma(w)^2\D w \nonumber \\
%=&-G^{-1}l^3\int_{0}^{1.37}(-G^T_{~T}) \sigma(w)^2\D w \nonumber \\
%\simeq& -2.65lG^{-1}=-2.65\sqrt{\hbar/G}=-2.65m_{\rm p} \nonumber \\
%\simeq& -5.76640\times 10^{-5}{\rm g}.
%\end{align} 

Negativity of $\lim_{x\to\infty}E(x)$ does not contradict the vanishing ADM mass because the integral of energy density only contributes to a part of the ADM mass.
As an expression of mass/energy which converges to the ADM mass at spacelike infinity~\cite{Hayward:1994bu}, the Misner-Sharp quasi-local mass $m_{\rm MS}$ is available~\cite{Misner:1964je}.
It is defined by 
\begin{align}
m_{\rm MS} := \frac{r}{2G}\left\{1-g^{AB}(D_A r)(D_B r)\right\}\label{MS-mass}
\end{align} 
for the following most general spherically symmetric spacetime;
\begin{align}
\D s^2 =g_{AB}(y)\D y^A\D y^B +r^2(y)(\D\theta^2+\sin^2\theta\D\phi^2),
\label{eq:ansatz}
\end{align} 
where $y^A~(A=0,1)$ are coordinates in a two-dimensional Lorentzian spacetime $(M^2, g_{AB})$ and $D_A$ is the covariant derivative on $(M^2, g_{AB})$.
For the spacetime (\ref{semi-vacuum}), or equivalently (\ref{semi-vacuum-w}), Eq.~(\ref{MS-mass}) gives
\begin{align}
m_{\rm MS} = \frac{r(x)}{2G}\biggl\{1-\biggl(\frac{\D r}{\D x}\biggl)^2\biggl\}=\frac{l\sigma}{2G}(1-{\sigma'}^2),\label{MS}
\end{align} 
which satisfies
\begin{align}
m_{\rm MS}'=4\pi l^3\mu \sigma^{2}\sigma'.\label{dMS}
\end{align} 

\begin{figure}[htbp]
\begin{center}
%\rotatebox{-90}{
\includegraphics[width=0.57\linewidth]{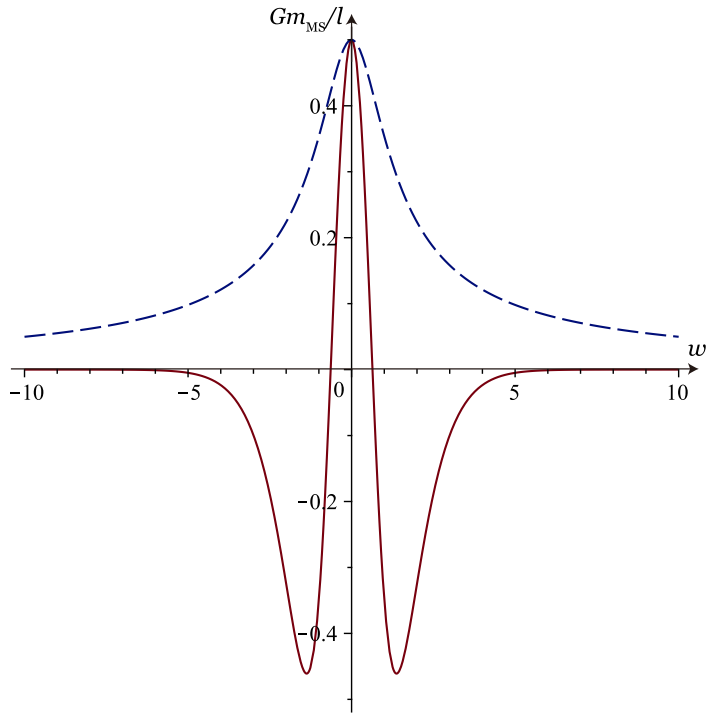}
%\subfigure[]{\includegraphics[width=0.7\linewidth]{Roberts-lambda1.eps}}
%\subfigure[]{\includegraphics[width=0.7\linewidth]{Roberts-lambda2.eps}}
%}
\caption{\label{MS-mass-wormhole} The Misner-Sharp mass (\ref{MS}) as a function of $w(=x/l)$ for the wormhole (\ref{semi-vacuum-w}) (solid) and the massless Ellis-Bronnikov wormhole $\sigma(w)=\sqrt{1+w^2}$ (dashed).
}
\end{center}
\end{figure}

In Fig.~\ref{MS-mass-wormhole}, we plot $m_{\rm MS}$ for our wormhole spacetime (\ref{semi-vacuum-w}) and the massless Ellis-Bronnikov wormhole with $\sigma(w)=\sqrt{1+w^2}$.
In both cases, $m_{\rm MS}$ takes the positive maximum value $m_{\rm MS}(0)=l/(2G)=m_{\rm p}/2\times (l/l_{\rm p})$ at the throat and converges to zero, the value of the ADM mass, as $w\to \pm \infty$.
By Eq.~(\ref{dMS}), the signs of $m_{\rm MS}'$ and $\sigma'$ are the same (opposite) in a region where $\mu>(<)0$ holds.
Thus, in our wormhole spacetime (\ref{semi-vacuum-w}), $m_{\rm MS}$ decreases as $|w|$ increases in the region of $|w|\lesssim 1.37$ and becomes zero at $w\simeq 0.639$.
After taking the negative minimum value at $m_{\rm MS}(|1.37|)\simeq -0.461lG^{-1}=-0.461m_{\rm p}\times (l/l_{\rm p})$ at $|w|\simeq 1.37$, $m_{\rm MS}$ starts to increase in the region of $|w|\gtrsim 1.37$ and rapidly converges to zero from below as $|w|\to \infty$.
This behavior is quite different from the massless Ellis-Bronnikov wormhole, for which $m_{\rm MS}$ is positive and monotonically decreasing, due to the negative energy everywhere, and converges to zero from above as $|w|\to \infty$\footnote{It might look counterintuitive that the quasi-local mass $m_{\rm MS}$ of the ``massless'' Ellis-Bronnikov wormhole is ``positive'' in the whole domain $|w|<\infty$ in spite of the ``negative'' energy density everywhere.
In fact, even a wormhole with positive ADM masses evaluated at both spacelike infinities is possible with non-positive energy density everywhere.
Such an example can be constructed by gluing two Schwarzschild spacetimes with the same positive ADM mass at a timelike hypersurface outside the horizon.
In this spacetime, the negative energy density is localized only at the throat.
Of course, these wormholes do not conflict with the positive mass theorem~\cite{Schon:1979rg,Schon:1981vd,Nester:1981bjx,Witten:1981mf} stating that, {\it under the dominant energy condition}, the ADM mass of an asymptotically flat and {\it regular} spacetime is non-negative and the spacetime is Minkowski if the ADM mass is zero.}

%------------<fig>---------------------------
%--------------<fig>-----------------------

The energy density (\ref{rho}) in our wormhole spacetime (\ref{semi-vacuum}) takes the minimum negative value $\mu(0)=-3/(8\pi Gl^2)$ at the throat $x=0$.
Let us estimate the value of $l$ under the assumption that $\mu(0)$ is comparable to the energy density of the Casimir effect in a laboratory, which is the best-known example of physical situations in which negative energy densities arise as a quantum effect.
Due to the Casimir effect, two ideally parallel conducting plates in vacuo experience an attractive force.
In the case of infinite plane plates separated by distance $L$ along the $z$-axis in the standard $(t,x,y,z)$ coordinates, one may deduce on symmetry grounds and dimensional considerations that the orthonormal components in the local Lorentz frame of the energy-momentum tensor of the electromagnetic field takes the following form~\cite{Brown:1969na,Fewster:2012yh,Martin-Moruno:2013wfa};
\begin{equation} 
\label{Casimir}
T^{(a)(b)}:=T^{\mu\nu}E_\mu^{(a)}E_\nu^{(b)}=\frac{\hbar \pi^2}{720 L^4}{\rm diag}(-1,1,1,-3).
\end{equation}
Hence, the energy density of the Casimir effect is given by $\mu_{\rm C}:=T^{(0)(0)}=-\hbar \pi^2/(720 L^4)$. 
Then, $|\mu_{\rm C}|\simeq |\mu(0)|$ gives
\begin{align}
%\frac{\hbar c\pi^2}{720 L^4}\ge \frac{3c^4}{8\pi G\alpha^2 l_{\rm p}^2}\\
%\frac{L^4}{l_{\rm p}^4}\le \frac{\alpha^2 \pi^3}{270 }.\\
\frac{l}{l_{\rm p}}\simeq\sqrt{\frac{270 }{\pi^3}}\frac{L^2}{l_{\rm p}^2}\simeq 2.95\times \frac{L^2}{l_{\rm p}^2}.
\end{align}
The Casimir effect has been confirmed by experiments with $L\simeq 10^{-7}$--$10^{-6}{\rm m}$~\cite{Klimchitskaya:2006rw,Klimchitskaya:2009cw}.
With this value of $L$, we obtain $l/l_{\rm p}\simeq 10^{56}$--$10^{58}$ and hence $l\simeq 10^{21}$--$10^{23}{\rm m}$.
This value of $l$ is extremely large because the Casimir effect in a laboratory is a quantum effect of matter, not of gravity.
$l$ should be close to $l_{\rm p}$ if the wormhole (\ref{semi-vacuum}) is realized in the semiclassical regime where quantum effects of gravity are taken into account.

\subsection{Axisymmetric model}
\label{sec:axial}

Now we generalize our spherically symmetric model (\ref{semi-vacuum}) to be axisymmetric inspired by the massless Kerr vacuum spacetime.
The Kerr vacuum spacetime is given in the Boyer-Lindquist coordinates $\{t,r,\theta,\phi\}$ as
\begin{align}
\D s^2=&-\biggl(1-\frac{2Mr}{r^2+a^2\cos^2\theta}\biggl)\D t^2-\frac{4aMr\sin^2\theta}{r^2+a^2\cos^2\theta}\D t\D\phi \nonumber\\
&+\frac{r^2+a^2\cos^2\theta}{r^2+a^2-2rM}\D r^2+(r^2+a^2\cos^2\theta)\D\theta^2+\biggl(r^2+a^2+\frac{2a^2Mr\sin^2\theta}{r^2+a^2\cos^2\theta}\biggl)\sin^2\theta\D\phi^2, \label{BL}
\end{align} 
where $M$ is the ADM mass and the constant $a$ characterizes the angular momentum of the spacetime.
It is well-known that $(r,\theta)=(0,\pi/2)$ is a ring-like curvature singularity for $M\ne 0$.
In the massless case $M=0$, in contrast, $(r,\theta)=(0,\pi/2)$ is not a curvature singularity but a ring-like conical singularity with angle excesses~\cite{Gibbons:2017djb}.

\begin{figure}[htbp]
\begin{center}
%\rotatebox{-90}{
\includegraphics[width=1.0\linewidth]{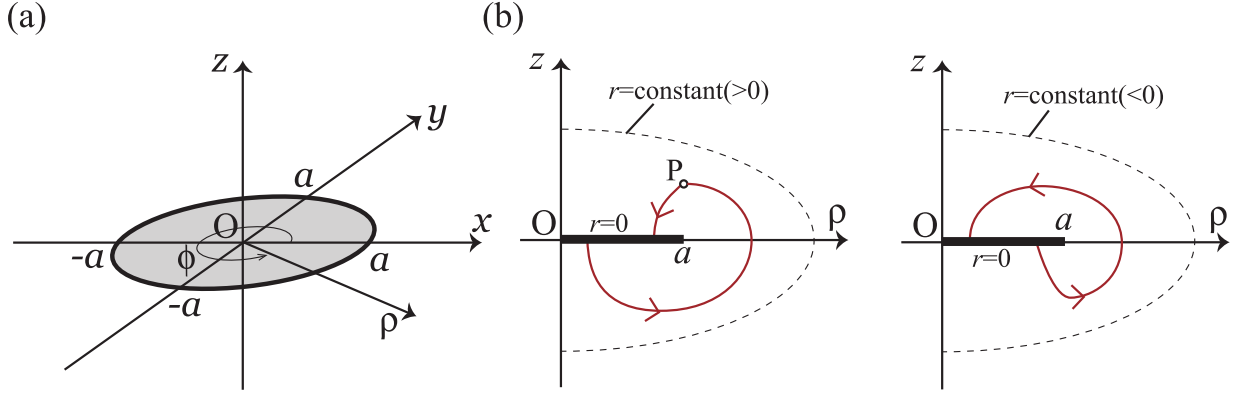}
%\subfigure[]{\includegraphics[width=0.7\linewidth]{Roberts-lambda1.eps}}
%\subfigure[]{\includegraphics[width=0.7\linewidth]{Roberts-lambda2.eps}}
%}
\caption{\label{MasslessKerr} (a) The Cartesian coordinates $(x,y,z)$ and the cylindrical coordinates $(\rho,\phi,z)$. (b) The cylindrical coordinates $(\rho,z)$ for the massless Kerr spacetime (\ref{BL-r=0}) with $r>0$ (left) and $r<0$ (right). A closed curve from a point P wraps around a ring $(\rho,z)=(a,0)$ by crossing twice a disk defined by $\rho\in[0,a)$ with $z=0$.
}
\end{center}
\end{figure}

The Kerr spacetime (\ref{BL}) with $M=0$ is the following flat spacetime in the oblate spheroidal coordinates:
\begin{align}
\D s^2=-\D t^2+\frac{r^2+a^2\cos^2\theta}{r^2+a^2}\D r^2+(r^2+a^2\cos^2\theta)\D\theta^2+(r^2+a^2)\sin^2\theta\D\phi^2, \label{BL-r=0}
\end{align}
which is obtained from the following cylindrical coordinates $\{t,\rho,\phi,z\}$
\begin{align}
\D s^2=-\D t^2+\D\rho^2+\rho^2\D\phi^2+\D z^2\label{cylinder}
\end{align}
by coordinate transformations $\rho=\sqrt{r^2+a^2}\sin\theta$ and $z=r\cos\theta$ satisfying
\begin{align}
\frac{\rho^2}{r^2+a^2}+\frac{z^2}{r^2}=1.\label{spheroid}
\end{align}
The domains of $\rho$ and $z$ are $0\le \rho<\infty$ and $-\infty<z<\infty$, respectively, and Eq.~(\ref{spheroid}) shows that $r=$constant represents a spheroid.
In the limit $r\to 0$, this spheroid reduces to a ``disk'' described by $\rho\in[0,a]$ with $z=0$, while $z=0$ outside this disk corresponds to $\theta=\pi/2$ with $r^2=\rho^2-a^2$. (See Fig.~\ref{MasslessKerr}(a).)
Since the metric (\ref{BL}) is of class $C^2$ at $r=0$ with $\theta\ne \pi/2$, the spacetime can be extended beyond the interior of the disk described by $\rho\in[0,a)$ with $z=0$.
Then, in order to remove the discontinuity at $z=0$, the extension of the spacetime requires attachment to a spacetime region with a different sign of $r$~\cite{Gibbons:2017djb}.
As a result, a closed curve wrapping around the ring $(\rho,z)=(a,0)$ crosses the disk twice. (See Fig.~\ref{MasslessKerr}(b).)
This shows that the ring $(\rho,z)=(a,0)$, or equivalently $(r,\theta)=(0,\pi/2)$, is a conical singularity with angle excesses and therefore the ring may be interpreted as a cosmic string with negative tension~\cite{Gibbons:2016bok}. (See also~\cite{Gibbons:2017jzk}.) 
Therefore, the massless Kerr spacetime (\ref{BL-r=0}) describes an axisymmetric static traversable wormhole with a conical singularity at $(r,\theta)=(0,\pi/2)$~\cite{Gibbons:2017djb}.
It is known that the distributional energy-momentum tensor density of a cosmic string along the $z$-axis in the isotropic (Cartesian) coordinate $\{t,x,y,z\}$ takes the following form;
\begin{equation} 
\label{T-string}
\widehat{\sqrt{-g}T^\mu_{~\nu}}=\nu\delta^{(2)}(x,y){\rm diag}(-1,0,0,-1),
\end{equation}
where $\delta^{(2)}(x,y)$ is the two-dimensional Dirac distribution~\cite{Anderson:2003gg}. (See Eqs.~(7.80) and (10.119) in the textbook~\cite{Anderson:2003gg} adopting the Minkowski signature $(+,-,-,-)$.)
According to Eqs.~(\ref{NEC})--(\ref{SEC}), the energy-momentum tensor for such a cosmic string with negative tension ($\nu<0$) violates all the standard energy condition.
%------------<fig>---------------------------
%--------------<fig>-----------------------

Based on this fact, we deform our spherically symmetric wormhole (\ref{semi-vacuum}) to be axisymmetric as
\begin{align}
\label{semi-vacuum-axi}
\begin{aligned}
\D s^2=&-\D t^2+\frac{\Sigma(x,\theta)}{\Delta(x)}\D x^2+\Sigma(x,\theta)\D\theta^2+\Delta(x)\sin^2\theta\D\phi^2, \\
&\Sigma(x,\theta):=r(x)^2+a^2\cos^2\theta,\qquad \Delta(x):=r(x)^2+a^2,
\end{aligned}
\end{align} 
where $r(x)$ is given in Eq.~(\ref{semi-vacuum}) and $a$ is now a deformation parameter.
The domain of $x$ is $-\infty<x<\infty$ and the spacetime is symmetric with respect to $x=0$.
The spacetime is asymptotically flat as $x\to \pm \infty$ with the vanishing ADM mass and uniformly converges to the spherically symmetric model (\ref{semi-vacuum}) in the limit $a\to 0$.
The spacetime is analytic everywhere and $\theta=0, \pi$ are just coordinate singularities.
Near a coordinate singularity $\theta=0$, the metric (\ref{semi-vacuum-axi}) becomes 
\begin{align}
\D s_{\theta=0}^2\simeq &-\D t^2+\D x^2+(r(x)^2+a^2)(\D\theta^2+\theta^2\D\phi^2).
\end{align} 
Since the two-dimensional angular part is in the flat Rindler form, there is no conical singularity at the pole $\theta=0$.
(One can show the same at $\theta=\pi$.)

\begin{figure}[htbp]
\begin{center}
%\rotatebox{-90}{
\includegraphics[width=0.7\linewidth]{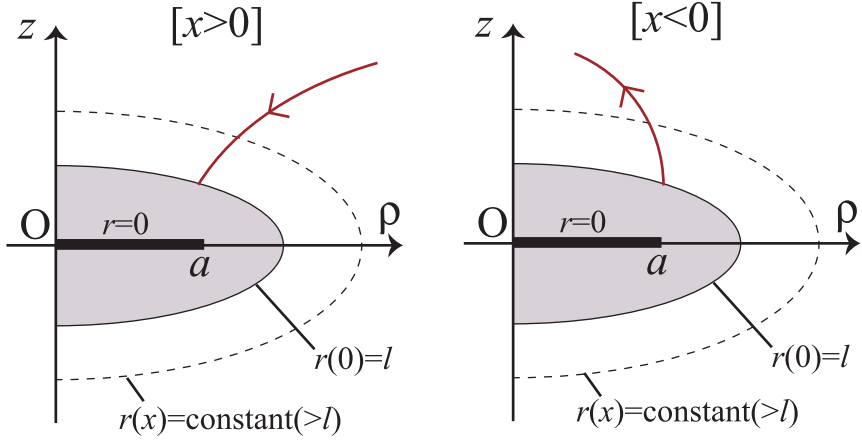}
%\subfigure[]{\includegraphics[width=0.7\linewidth]{Roberts-lambda1.eps}}
%\subfigure[]{\includegraphics[width=0.7\linewidth]{Roberts-lambda2.eps}}
%}
\caption{\label{WormholeThroat-Axial} The cylindrical variables $(\rho,z)$ for the axisymmetric wormhole spacetime (\ref{semi-vacuum-axi}). Domains with $x>0$ (left) and $x<0$ (right) are attached at the wormhole throat $x=0$, where $r(0)=l$ holds.
A solid curve passing the wormhole throat is drawn.
Shaded regions do not exist in the spacetime (\ref{semi-vacuum-axi}).
}
\end{center}
\end{figure}

For the spacetime (\ref{semi-vacuum-axi}), the surface area $A(x_0)$ with constant $x(=x_0)$ on a spacelike hypersurface with constant $t$ and its derivative are given by 
\begin{align}
&A(x_0)=2\pi \int_0^\pi \sqrt{{\Sigma}(x_0,\theta) {\Delta}(x_0)}\sin\theta\D\theta\\
\to~~&\frac{\D A}{\D x_0}=2\pi r(x_0)r'(x_0)\int_0^\pi\frac{\Sigma(x_0,\theta)+\Delta(x_0)}{\sqrt{{\Sigma}(x_0,\theta) {\Delta}(x_0)}}\sin\theta\D\theta.
\end{align} 
The location of a wormhole throat $x=x_{\rm t}$ is determined by $\D A/\D x_0|_{x_0=x_{\rm t}}=0$, or equivalently $r'(x_{\rm t})=0$.
Therefore, the throat is located at $x=0$ as in the spherically symmetric case, so that the axisymmetric spacetime (\ref{semi-vacuum-axi}) represents an asymptotically flat axisymmetric static wormhole.
The induced metric on the throat with constant $t$ is given by
\begin{align}
\D s_{x=0, \D t=0}^2=&(l^2+a^2\cos^2\theta)\D\theta^2+(l^2+a^2)\sin^2\theta\D\phi^2,\label{induced-throat}
\end{align} 
which is a deformed two-sphere for $a\ne 0$ with the Riemann tensor $R^{\theta\phi}_{~~~\theta\phi}=l^2/(l^2+a^2\cos^2\theta)^2$. 
Figure~\ref{WormholeThroat-Axial} shows the domains of the cylindrical variables $\rho=\sqrt{r(x)^2+a^2}\sin\theta$ and $z=r(x)\cos\theta$ for the wormhole spacetime (\ref{semi-vacuum-axi}) satisfying
\begin{align}
\frac{\rho^2}{r(x)^2+a^2}+\frac{z^2}{r(x)^2}=1.\label{spheroid2}
\end{align}
Since the metric (\ref{semi-vacuum-axi}) is analytic everywhere and reduces to the massless Kerr spacetime (\ref{BL-r=0}) as $l\to 0$ (and then $r(x)\to x$), the wormhole (\ref{semi-vacuum-axi}) may be interpreted as a regularized version of the massless Kerr wormhole (\ref{BL-r=0}).
%------------<fig>---------------------------
%--------------<fig>-----------------------

Now we show that an exotic matter in the wormhole spacetime (\ref{semi-vacuum-axi}) is localized around the throat $x=0$.
With dimensionless coordinates $T:=t/l$ and $w:=x/l$, the spacetime (\ref{semi-vacuum-axi}) is written as
\begin{align}
\label{w-metric}
\begin{aligned}
\D s^2&=l^2\biggl[-\D T^2+\frac{{\bar \Sigma}(w,\theta)}{{\bar \Delta}(w)}\D w^2+{\bar \Sigma}(w,\theta)\D\theta^2+{\bar \Delta}(w)\sin^2\theta\D\phi^2\biggl], \\
&{\bar \Sigma}(w,\theta):={\Sigma}(x(w),\theta)/l^2=\sigma(w)^2+{\bar a}^2\cos^2\theta,\\
&{\bar \Delta}(w):={\Delta}(x(w))/l^2=\sigma(w)^2+{\bar a}^2,
\end{aligned}
\end{align} 
where $\sigma(w)$ is defined in Eq.~(\ref{semi-vacuum-w}) and ${\bar a}:=a/l$ is a dimensionless deformation parameter.
The corresponding energy-momentum tensor is diagonal such that $T^\mu_{~\nu}:=(8\pi G)^{-1}G^\mu_{~\nu}={\rm diag}(-\mu,p_1,p_2,p_3)$, where
\begin{align}
&\mu=\frac{1}{8\pi Gl^2}\biggl[-\frac{({\bar \Sigma}+{\bar \Delta})(\sigma\sigma''+{\sigma'}^2-1)}{{\bar\Sigma}^2}+\frac{\sigma^2({\sigma'}^2-1)({\bar \Delta}+{\bar a}^2\sin^2\theta)}{{\bar\Sigma}^3}\biggl],\label{rho-axi}\\
&p_1=\frac{\sigma^2({\sigma'}^2-1)}{8\pi Gl^2{\bar\Sigma}^2},\qquad p_2=\frac{1}{8\pi Gl^2}\biggl[\frac{\sigma\sigma''}{{\bar\Sigma}}+\frac{({\sigma'}^2-1){\bar a}^2\cos^2\theta}{{\bar\Sigma}^2}\biggl],\label{p2-axi}\\
&p_3=\frac{1}{8\pi Gl^2}\biggl[\frac{\sigma\sigma''{\bar \Delta}}{{\bar\Sigma}^2}-\frac{{\bar a}^2({\sigma'}^2-1)\{\sigma^2-(2\sigma^2+{\bar a}^2)\cos^2\theta\}}{{\bar\Sigma}^3}\biggl].\label{p3-axi}
\end{align}
Hence, the spacetime (\ref{semi-vacuum-axi}), or equivalently the spacetime (\ref{w-metric}), is an exact solution in general relativity with an anisotropic fluid, of which energy-momentum tensor is given by 
\begin{align}
{T}_{\mu\nu}=&\mu u_\mu u_\nu+S_{\mu\nu}.\label{T-fluid-axi}
\end{align} 
Here $S_{\mu\nu}$ is the stress tensor satisfying the following eigenvalue equations~\cite{Pimentel:2016jlm}:
\begin{align}
S_{\mu\nu}s_{(1)}^\nu=p_1 g_{\mu\nu}s_{(1)}^\nu,\qquad S_{\mu\nu}s_{(2)}^\nu=p_2 g_{\mu\nu}s_{(2)}^\nu,\qquad S_{\mu\nu}s_{(3)}^\nu=p_3 g_{\mu\nu}s_{(3)}^\nu, \label{eigen-axi}
\end{align} 
where $s_{(i)}^\mu~(i=1,2,3)$ are distinct unit orthogonal spacelike vectors satisfying $u_\mu s_{(i)}^\mu=0$.
In the spacetime (\ref{semi-vacuum-axi}), $u^\mu$ and $s_{(i)}^\mu~(i=1,2,3)$ are given by 
\begin{align}
&u^\mu\frac{\partial}{\partial x^\mu}=\frac{1}{l}\frac{\partial}{\partial x^\mu},\qquad s_{(1)}^\mu\frac{\partial}{\partial x^\mu}=\frac{1}{l}\sqrt{\frac{{\bar \Delta}}{{\bar \Sigma}}}\frac{\partial}{\partial x^\mu},\\
&s_{(2)}^\mu\frac{\partial}{\partial x^\mu}=\frac{1}{l\sqrt{{\bar \Sigma}}}\frac{\partial}{\partial x^\mu},\qquad s_{(3)}^\mu\frac{\partial}{\partial x^\mu}=\frac{1}{l\sqrt{{\bar \Delta}}\sin\theta}\frac{\partial}{\partial x^\mu}.
\end{align}

Equations~(\ref{rho-axi})--(\ref{p3-axi}) give $\mu+p_1+p_2+p_3=0$ and 
\begin{align}
\mu+p_1=&\frac{1}{8\pi Gl^2}\biggl[-\frac{({\bar \Sigma}+{\bar \Delta})(\sigma\sigma''+{\sigma'}^2-1)}{{\bar\Sigma}^2}+\frac{2\sigma^2({\sigma'}^2-1){\bar \Delta}}{{\bar\Sigma}^3}\biggl],\\
\mu+p_2=&\frac{1}{8\pi Gl^2}\biggl[-\frac{{\bar \Delta}(\sigma\sigma''+{\sigma'}^2-1)}{{\bar\Sigma}^2}+\frac{2\sigma^2({\sigma'}^2-1){\bar a}^2\sin^2\theta}{{\bar\Sigma}^3}\biggl], \\
\mu+p_3=&-\frac{\sigma\sigma''+{\sigma'}^2-1}{8\pi Gl^2{\bar\Sigma}},\label{rho+p3-axi}\\
%%%%%%%%%%%%%%%%%%%%
\mu-p_1=&\frac{1}{8\pi Gl^2}\biggl[-\frac{({\bar \Sigma}+{\bar \Delta})(\sigma\sigma''+{\sigma'}^2-1)}{{\bar\Sigma}^2}+\frac{2\sigma^2({\sigma'}^2-1){\bar a}^2\sin^2\theta}{{\bar\Sigma}^3}\biggl],\\
\mu-p_2=&\frac{1}{8\pi Gl^2}\biggl[-\frac{(2{\bar \Sigma}+{\bar \Delta})(\sigma\sigma''+{\sigma'}^2-1)}{{\bar\Sigma}^2}+\frac{2\sigma^2({\sigma'}^2-1){\bar \Delta}}{{\bar\Sigma}^3}\biggl], \\
\mu-p_3=&\frac{1}{8\pi Gl^2}\biggl[-\frac{({\bar \Sigma}+2{\bar \Delta})(\sigma\sigma''+{\sigma'}^2-1)}{{\bar\Sigma}^2}+\frac{2\sigma^2({\sigma'}^2-1)({\bar \Delta}+{\bar a}^2\sin^2\theta)}{{\bar\Sigma}^3}\biggl].
\end{align}
According to Eqs.~(\ref{NEC})--(\ref{SEC}), all the standard energy conditions are satisfied in the region where $\sigma\sigma''+{\sigma'}^2-1\le 0$ and ${\sigma'}^2-1\ge 0$ hold.
With $\sigma(w)$ given in Eq.~(\ref{semi-vacuum-w}), these inequalities are satisfied in the domain $|w|\gtrsim 1.60$.
On the other hand, by Eq.~(\ref{rho+p3-axi}), all the standard energy conditions are violated in the region where $\sigma\sigma''+{\sigma'}^2-1> 0$ holds.
This inequality corresponds to the domain $|w|\lesssim 1.60$ including the wormhole throat $w=0$, where $\sigma'=0$, $\sigma''=2$, and $\sigma=1$ hold.
Therefore, as in the spherically symmetric case, an exotic matter in the spacetime (\ref{semi-vacuum-axi}) is localized only near the wormhole throat.
We note that, in the case with $\sigma(w)=\sqrt{1+w^2}$ corresponding to the massless Ellis-Bronnikov wormhole~\cite{Ellis1973,Bronnikov1973} in the spherically symmetric case, $\sigma\sigma''+{\sigma'}^2-1=0$ and ${\sigma'}^2-1=-1/(1+w^2)<0$ hold and hence all the standard energy conditions are violated everywhere.

%======================================%
%<<<<<<<<<<<< SECTION 1 >>>>>>>>>>>>>>%
%======================================%
\section{Geometric properties}
\label{app:Petrov}

In this section, we investigate geometric properties of the axisymmetric wormhole spacetime (\ref{semi-vacuum-axi}) and its spherically symmetric limit (\ref{semi-vacuum}).

\subsection{Petrov type}

First we clarify the Petrov type of the spacetime (\ref{semi-vacuum-axi}) by the method presented in Sec.~9 in the textbook~\cite{Chandrasekhar:1985kt}.
We introduce a complex null tetrad $\{l^\mu,n^\mu,m^\mu,{\bar m}^\mu\}$ as
\begin{align}
&l_\mu\D x^\mu=\frac{1}{\sqrt{2}}\biggl(-\D t-\sqrt{\Delta(x)}\sin\theta\D\phi\biggl),\\
&n_\mu\D x^\mu=\frac{1}{\sqrt{2}}\biggl(-\D t+\sqrt{\Delta(x)}\sin\theta\D\phi\biggl),\\
&m_\mu\D x^\mu=\frac{1}{\sqrt{2}}\biggl(\sqrt{\frac{\Sigma(x,\theta)}{\Delta(x)}}\D x+i\sqrt{\Sigma(x,\theta)}\D\theta\biggl),\\
&{\bar m}_\mu\D x^\mu=\frac{1}{\sqrt{2}}\biggl(\sqrt{\frac{\Sigma(x,\theta)}{\Delta(x)}}\D x-i\sqrt{\Sigma(x,\theta)}\D\theta\biggl),
\end{align} 
which satisfy $l_\mu n^\mu=-1$ and $m_\mu {\bar m}^\mu=1$.
Then, the Weyl scalars are computed to give
\begin{align}
\Psi_0 :=& C_{\alpha\beta\gamma\delta} l^\alpha m^\beta l^\gamma m^\delta=-\frac{rr''\Sigma-({r'}^2-1)(r^2-a^2\cos^2\theta)}{4\Sigma^2}\ , \label{Psi0}\\
\Psi_1 :=& C_{\alpha\beta\gamma\delta} l^\alpha n^\beta l^\gamma m^\delta=0\ , \\
\Psi_2 :=& C_{\alpha\beta\gamma\delta} l^\alpha m^\beta \bar{m}^\gamma n^\delta=\frac{(r^2+a^2+a^2\sin^2\theta)[rr''\Sigma-({r'}^2-1)(r^2-a^2\cos^2\theta)]}{12\Sigma^3}\ , \\
\Psi_3 :=& C_{\alpha\beta\gamma\delta} l^\alpha n^\beta \bar{m}^\gamma n^\delta=0\ , \\
\Psi_4 :=& C_{\alpha\beta\gamma\delta} n^\alpha \bar{m}^\beta n^\gamma \bar{m}^\delta=-\frac{rr''\Sigma-({r'}^2-1)(r^2-a^2\cos^2\theta)}{4\Sigma^2}(=\Psi_0).\label{Psi4}
\end{align} 

\begin{figure}[htbp]
\begin{center}
%\rotatebox{-90}{
\includegraphics[width=0.5\linewidth]{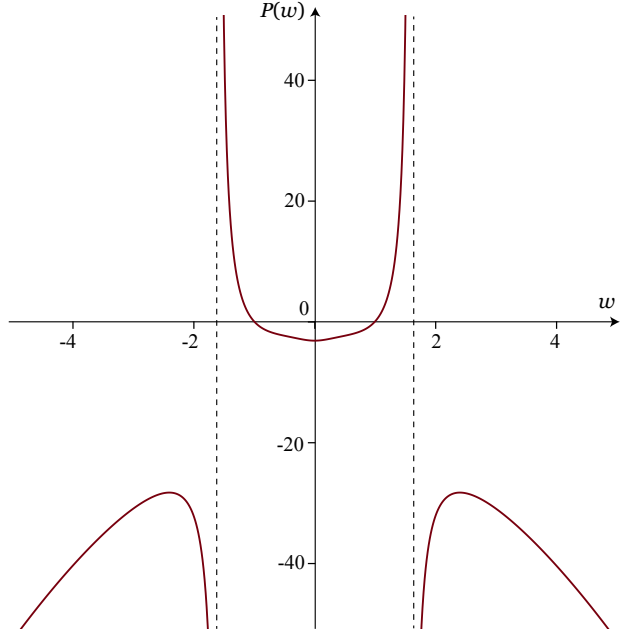}
%\subfigure[]{\includegraphics[width=0.7\linewidth]{Roberts-lambda1.eps}}
%\subfigure[]{\includegraphics[width=0.7\linewidth]{Roberts-lambda2.eps}}
%}
\caption{\label{PetrovOfunction} The function $P(w)$ in Eq.~(\ref{def-P}).
}
\end{center}
\end{figure}

In a spacetime region where $rr''\Sigma-({r'}^2-1)(r^2-a^2\cos^2\theta)= 0$ holds, all the Weyl scalars are zero, so that the spacetime is of Petrov type O there.
With $r(x)$ given in Eq.~(\ref{semi-vacuum}), this condition is written as
\begin{align}
\label{def-P}
\frac{a^2}{l^2}\cos^2\theta=-\frac{(w^2\tanh^2w + 2w\tanh w + w^2 - 3)(1+w\tanh w)^2}{3w^2\tanh^2 w - 2w\tanh w - w^2 - 1}:=P(w),
\end{align} 
where $w:=x/l$.
The function $P(w)$ is drawn in Fig.~\ref{PetrovOfunction}.
Since $0\le (a^2/l^2)\cos^2\theta\le a^2/l^2$ holds, Eq.~(\ref{def-P}) represents two disconnected closed two-surfaces which are located in domains of $x>0$ and $x<0$, respectively.
%------------<fig>---------------------------
%--------------<fig>-----------------------

In a spacetime region where $rr''\Sigma-({r'}^2-1)(r^2-a^2\cos^2\theta)\ne 0$ holds, we consider an invariant algebraic equation $\Psi_4Z^4+4\Psi_3 Z^3+6\Psi_2Z^2+4\Psi_1Z+\Psi_0=0$, which reduces to
\begin{align}
\Sigma Z^4-2(r^2+a^2+a^2\sin^2\theta)Z^2+\Sigma=0
\end{align} 
with Eqs.~(\ref{Psi0})--(\ref{Psi4}).
This algebraic equation admits solutions $Z=Z_1, Z_2, Z_3, Z_4$, where
\begin{align}
Z_1=&\frac{|\sqrt{r^2+a^2}+ a\sin\theta|}{\sqrt{r^2+a^2\cos^2\theta}},\qquad Z_2=-\frac{|\sqrt{r^2+a^2}+ a\sin\theta|}{\sqrt{r^2+a^2\cos^2\theta}},\\
Z_3=&\frac{|\sqrt{r^2+a^2}- a\sin\theta|}{\sqrt{r^2+a^2\cos^2\theta}},\qquad Z_4=-\frac{|\sqrt{r^2+a^2}- a\sin\theta|}{\sqrt{r^2+a^2\cos^2\theta}}.
\end{align} 
For $a\ne 0$, all these roots are different except at coordinate singularities $\theta=0,\pi$, so that the spacetime is of Petrov type I.
In the spherically symmetric case ($a=0$), we have $1=Z_1= Z_3\ne Z_2= Z_4=-1$ and therefore the spacetime is of Petrov type D.

\subsection{Geodesic analysis}
\label{sec:geodesics}

Next we study geodesics in the axisymmetric wormhole spacetime (\ref{semi-vacuum-axi}).
Let $\gamma$ be an affinely parametrized geodesic represented as $x^\mu=(t(\lambda),x(\lambda),\theta(\lambda),\phi(\lambda))$, where $\lambda$ is an affine parameter. 
Let $k^\mu=({\dot t}, {\dot x}, {\dot \theta},{\dot \phi})$ be the tangent vector of $\gamma$, where a dot denotes differentiation with respect to $\lambda$.
The components of $k^\mu$ satisfy
\begin{align}
\varepsilon=&-{\dot t}^2+\frac{{\Sigma}(x,\theta)}{{\Delta}(x)}{\dot x}^2+{\Sigma}(x,\theta){\dot \theta}^2+{\Delta}(x)\sin^2\theta{\dot \phi}^2,\label{ds-gamma}
\end{align}
where $\varepsilon=-1,0,1$ correspond to $\gamma$ being timelike, null, and spacelike, respectively.
Since the spacetime (\ref{semi-vacuum-axi}) admits Killing vectors $\xi^\mu(\partial/\partial x^\mu)=\partial/\partial t$ and $\Phi^\mu(\partial/\partial x^\mu)=\partial/\partial \phi$, $E:=-\xi_\mu k^\mu={\dot t}$ and $L:=\Phi_\mu k^\mu={\Delta}\sin^2\theta{\dot \phi}$ are conserved along $\gamma$, which represent the energy and angular momentum of the geodesic particle, respectively.
With $E$ and $L$, we rewrite Eq.~(\ref{ds-gamma}) as
\begin{align}
\varepsilon=&-E^2+\frac{{\Sigma}}{{\Delta}}{\dot x}^2+{\Sigma}{\dot \theta}^2+\frac{L^2}{{\Delta}\sin^2\theta}.\label{geodesic-epsilon}
\end{align}

In addition, since the spacetime (\ref{semi-vacuum-axi}) admits a two-rank Killing tensor $K^{\mu\nu}$ satisfying $\nabla_{(\rho}K_{\mu\nu)}=0$, $C:=K_{\mu\nu}k^\mu k^\nu$ is constant along $\gamma$.
In the spacetime (\ref{semi-vacuum-axi}), $K^{\mu\nu}$ is given by 
\begin{align}
K^{\mu\nu}=\Sigma(\eta^\mu \zeta^\nu+\zeta^\mu \eta^\nu )+r(x)^2 g^{\mu\nu},
\end{align}
where $\eta^\mu$ and $\zeta^\mu$ are null vectors defined by 
\begin{align}
\eta^\mu=\biggl(1,1,0,\frac{a}{\Delta}\biggl),\qquad \zeta^\mu=\biggl(\frac{\Delta}{2\Sigma},-\frac{\Delta}{2\Sigma},0,\frac{a}{2\Sigma}\biggl).
\end{align}
With the following non-zero components of $K_{\mu\nu}$;
\begin{align}
&K_{tt}=a^2,\qquad K_{t\phi}(=K_{\phi t})=-\Delta a\sin^2\theta,\\
&K_{xx}=-\frac{\Sigma a^2\cos^2\theta}{\Delta},\qquad K_{\theta\theta}=r(x)^2\Sigma,\\
&K_{\phi\phi}=\Delta \sin^2\theta(r(x)^2+a^2\sin^2\theta),
\end{align}
we obtain
\begin{align}
C=&{ a}^2E^2-2{ a}EL-\frac{{ \Sigma} { a}^2\cos^2\theta}{{ \Delta}}{\dot x}^2+{ \Sigma}r(x)^2{\dot\theta}^2+\frac{L^2(r(x)^2+a^2\sin^2\theta)}{\Delta \sin^2\theta}.\label{geodesic-C}
\end{align}
Finally, Eqs~(\ref{geodesic-epsilon}) and (\ref{geodesic-C}) give the following set of ordinary differential equations to determine $x(\lambda)$ and $\theta(\lambda)$ along $\gamma$:
\begin{align}
{\dot x}^2=&\frac{\Delta[ (E^2 + \varepsilon) r(x)^2-(C- a^2E^2+ 2aEL) ]+a^2L^2}{{\Sigma}^2},\label{g-master-x}\\
{\dot\theta}^2=&\frac{-(E^2 + \varepsilon)a^2\sin^4\theta+( \varepsilon a^2+C+ 2aEL)\sin^2\theta-L^2}{\Sigma^2\sin^2\theta}.\label{g-master-theta}
\end{align}
A geodesic exists in a spacetime region where the numerators in the right-hand sides of Eqs.~(\ref{g-master-x}) and (\ref{g-master-theta}) are both non-negative.
%------------<fig>---------------------------
\begin{figure}[htbp]
\begin{center}
%\rotatebox{-90}{
\includegraphics[width=1.0\linewidth]{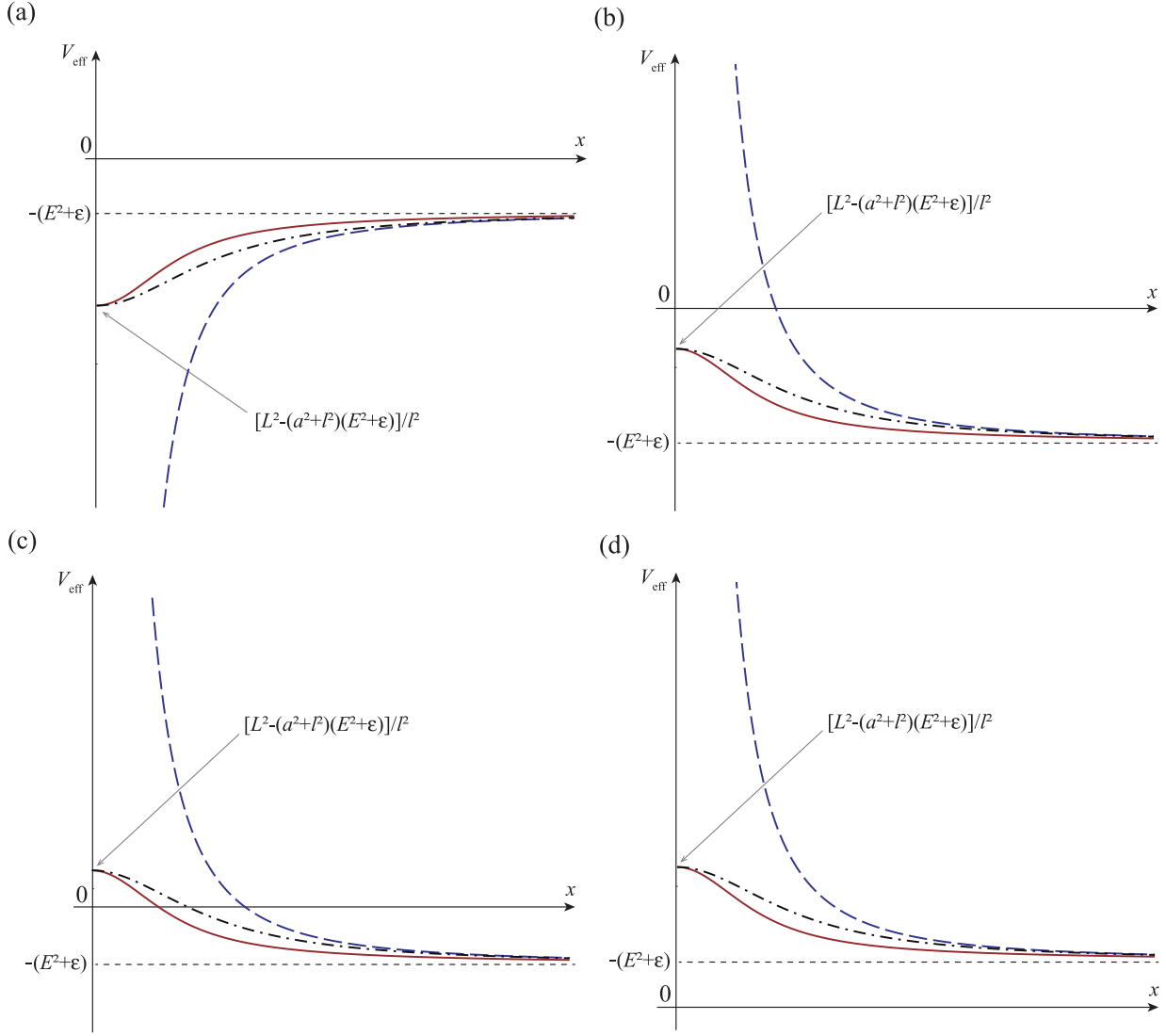}
%\subfigure[]{\includegraphics[width=0.7\linewidth]{Roberts-lambda1.eps}}
%\subfigure[]{\includegraphics[width=0.7\linewidth]{Roberts-lambda2.eps}}
%}
\caption{\label{Wormhole-geodesic-axi} Effective potentials $V_{\rm eff}(x)$ given by Eq.~(\ref{ODE-V}) for geodesics on the equatorial plane $\theta=\pi/2$ in the domain $x\in[0,\infty)$ with $r(x)=l+x\tanh(x/l)$ (solid), $r(x)=\sqrt{x^2+l^2}$ (dashdotted), and $r(x)=x$ (dashed) in the cases of (a) $E^2 + \varepsilon>0$ and $L^2<a^2(E^2 + \varepsilon)$, (b) $E^2 + \varepsilon>0$ and $a^2(E^2 + \varepsilon)<L^2<(a^2+l^2)(E^2 + \varepsilon)$, (c) $E^2 + \varepsilon>0$ and $L^2>(a^2+l^2)(E^2 + \varepsilon)$, and (d) $E^2 + \varepsilon<0$. 
}
%\caption{\label{Wormhole-geodesic-axi} Effective potentials $V_{\rm eff}(x)$ given by Eq.~(\ref{ODE-V}) for geodesics on the equatorial plane $\theta=\pi/2$ in the domain $x\in[0,\infty)$ of the axisymmetric wormhole spacetime (\ref{semi-vacuum-axi}) (solid) and massless Kerr spacetime (\ref{BL-r=0}) (dashed) with $L^2>a^2(E^2 + \varepsilon)$ (left) and $L^2<a^2(E^2 + \varepsilon)$ (right). 
%}
\end{center}
\end{figure}
%--------------<fig>-----------------------

Hereafter we focus on geodesics confined on the equatorial plane $\theta=\pi/2$ for simplicity.
Along such geodesics, $C= (aE-L)^2$ is satisfied by Eq.~(\ref{g-master-theta}) and then Eq.~(\ref{g-master-x}) reduces to
\begin{align}
&{\dot x}^2+V_{\rm eff}(x)=0,\label{ODE-r-0}\\
&V_{\rm eff}(x):=-(E^2 + \varepsilon)+\frac{L^2-a^2(E^2 + \varepsilon)}{r(x)^2}.\label{ODE-V}
\end{align}
The effective potential (\ref{ODE-V}) is symmetric with respect to $x=0$.
Since ${\dot x}^2=-V_{\rm eff}(x)\ge 0$ is required, an inequality $E^2+\varepsilon> 0$ must be satisfied for a geodesic coming from an asymptotically flat region ($r(x)\to \infty$).
In Fig.~\ref{Wormhole-geodesic-axi}, we compare the forms of $V_{\rm eff}(x)$ given by Eq.~(\ref{ODE-V}) with three different functions of $r(x)$:
\begin{enumerate}
\item $r(x)=l+x\tanh(x/l)$ for our axisymmetric wormhole (\ref{semi-vacuum-axi}).
\item $r(x)=\sqrt{x^2+l^2}$ corresponding to the massless Ellis-Bronnikov wormhole~\cite{Ellis1973,Bronnikov1973} in the spherically symmetric case ($a=0$).
\item $r(x)=x$ for the massless Kerr wormhole (\ref{BL-r=0}).
\end{enumerate}
While an exotic matter is localized near the wormhole throat in our wormhole spacetime (\ref{semi-vacuum-axi}) with $r(x)$ given in Eq.~(\ref{semi-vacuum}), all the standard energy conditions are violated everywhere in the wormhole spacetime with $r(x)=\sqrt{x^2+l^2}$.
Nevertheless, Fig.~\ref{Wormhole-geodesic-axi} shows that there is no qualitative difference of the geodesic behaviors in these two spacetimes.
This indicates that the behaviors of geodesic in a wormhole spacetime do not depend so much on whether an exotic matter is localized or not.

In these two wormhole spacetimes, $V_{\rm eff}(x)$ takes an extreme value $V_{\rm eff}(0)=[L^2-(a^2+l^2)(E^2 + \varepsilon)]/l^2$ at the wormhole throat $x=0$.
Since $\D V_{\rm eff}/\D x=0$ holds at $x=0$, a geodesic satisfying $L^2=(a^2+l^2)(\varepsilon+E^2)$ can stay at the wormhole throat $x=0$ on the equatorial plane $\theta=\pi/2$, which shows that the wormhole throat is a photon sphere (for $\varepsilon=0$).
As shown in Fig.~\ref{Wormhole-geodesic-axi}(d), there is no geodesic on the equatorial plane with $E^2 + \varepsilon\le 0$.
In the case of $E^2 + \varepsilon> 0$, as shown in Fig.~\ref{Wormhole-geodesic-axi}(a)--(c), a geodesic coming from an asymptotically flat region $x\to \infty$ can pass through the wormhole throat $x=0$ if $L^2<(a^2+l^2)(E^2 + \varepsilon)$ is satisfied.
In contrast, a geodesic with $E^2 + \varepsilon> 0$ and $L^2\ge (a^2+l^2)(E^2 + \varepsilon)(>a^2(E^2 + \varepsilon))$ cannot arrive the wormhole throat.
Hence, if the deformation parameter $a$ in the wormhole spacetime (\ref{semi-vacuum-axi}) is larger, geodesics with a given energy $E$ can reach the throat $x=0$ even with a larger angular momentum $L$.

Now we consider the spherically symmetric wormhole (\ref{semi-vacuum}) which corresponds to the spacetime (\ref{semi-vacuum-axi}) with $a=0$.
As shown in Sec.~\ref{sec:spherical}, the total amount of the negative energy $E_{-}(l)$ given by Eq.~(\ref{negative-E}) supporting the wormhole can be arbitrarily small in the Minkowski limit $l\to 0$.
However, violation of the ANCC does not become arbitrarily small in this limit.
We can see this fact along a radial null geodesic $\gamma$.
Equation~(\ref{ODE-r-0}) with $a=0$ and $L=\varepsilon=0$ gives ${\dot x}=\pm |E|$ along such a geodesic, with which we compute
\begin{align}
\int_{-\infty}^\infty R_{\mu\nu}k^\mu k^\mu\D\lambda=&-\int_{-\infty}^\infty\frac{4|E|(1-w\tanh w)}{l\cosh^2w(1+w\tanh w )}\D w \nonumber\\
\simeq&-3.91l^{-1}|E|.\label{Rkk-result}
\end{align} 
As the integral (\ref{Rkk-result}) diverges as $l\to 0$, violation of the ANCC cannot be arbitrarily small in the Minkowski limit.
This peculiar property stems from the fact that the convergence of the spacetime (\ref{semi-vacuum}) to Minkowski is non-uniform.

\section{Summary and future prospects}
\label{sec:summary}

In this paper, we have first presented a simple static and spherically symmetric spacetime (\ref{semi-vacuum}) with vanishing ADM mass which describes a traversable wormhole characterized by a length parameter $l$ and reduces to Minkowski in a non-uniform manner as $l\to 0$.
The wormhole connects two distinct asymptotically flat regions ($x\to \pm\infty$) and the areal radius of its throat is exactly $l$.
In this spacetime, an exotic matter violating the standard energy conditions is localized near the wormhole throat and all the standard energy conditions are respected outside the proper radial distance approximately $1.60l$ from the throat.
If $l$ is identical to the Planck length $l_{\rm p}$, the curvature at the throat exceeds the Planck scale and a quantum gravity description should be necessary.
With $l\simeq 10l_{\rm p}$, in contrast, the curvature at the throat is sub-Planckian and then the spacetime (\ref{semi-vacuum}) may be a semi-classical model of traversable wormholes.
In this case, the total amount of the negative energy supporting the wormhole is only $E\simeq -26.5m_{\rm p}c^2$, which is the rest mass energy of about $-5.77\times 10^{-4}{\rm g}$.

Subsequently, we have generalized our spherically symmetric model (\ref{semi-vacuum}) to be axisymmetric with an additional deformation parameter $a$, which reduces to the massless Kerr vacuum wormhole as $l\to 0$.
Our axisymmetric wormhole spacetime (\ref{semi-vacuum-axi}) is analytic everywhere and regularizes a ring-like conical singularity of the massless Kerr wormhole by virtue of a localized exotic matter violating the standard energy conditions only near the wormhole throat.
The spacetime (\ref{semi-vacuum-axi}) with $a\ne 0$ is of Petrov type I in general and of Petrov type O at two disconnected closed two-surfaces.

Lastly, we have investigated the behavior of geodesics confined on the equatorial plane $\theta=\pi/2$ in our axisymmetric wormhole spacetime (\ref{semi-vacuum-axi}).
As the deformation parameter $a$ increases, the geodesics with a given energy $E$ can reach the wormhole throat even with larger angular momentum $L$.
We have shown that the behavior of geodesics in our wormhole spacetime is qualitatively similar to that in a wormhole spacetime in which all the standard energy conditions are violated everywhere.
This suggests that the geodesic behavior does not qualitatively change by the localization of an exotic matter.

It goes without saying that dynamical stability is the most important problem to single out a realistic traversable wormhole spacetime.
Since the region where the energy conditions are violated is tiny with small $l$, our wormhole spacetime (\ref{semi-vacuum}) or (\ref{semi-vacuum-axi}) could be a dynamically stable configuration.
While the spacetime (\ref{semi-vacuum}) and (\ref{semi-vacuum-axi}) are solutions in general relativity with an anisotropic fluid, it has been reported that beyond-Horndeski theories~\cite{Zumalacarregui:2013pma,Gleyzes:2014dya} could admit dynamically stable static wormholes~\cite{Franciolini:2018aad,Mironov:2018uou}.
To identify theories of gravity which admit the spacetime (\ref{semi-vacuum}) or (\ref{semi-vacuum-axi}) as a solution and clarify its dynamical stability are important tasks.
We leave these problems for future investigations.

\subsection*{Acknowledgements}
The author thanks Hajime Sotani for helpful comments.

%\appendix

\end{document}